\documentclass[twocolumn,aps,showpacs,prl]{revtex4}
\usepackage{epsfig}
\usepackage[english]{babel}
\usepackage{latexsym}
\usepackage{graphics}
\usepackage{subfigure}
\usepackage{epsfig}
\usepackage{graphicx}
\usepackage{dcolumn}
\usepackage{amsmath}

\newcommand{\be}{\begin{equation}}
\newcommand{\ee}{\end{equation}}
\newcommand{\bey}{\begin{eqnarray}}
\newcommand{\eey}{\end{eqnarray}}
\newcommand{\bw}{\begin{widetext}}
\newcommand{\ew}{\end{widetext}}

\begin{document}

\title{Singularity of Berry Connections   Inhibits the Accuracy of Adiabatic Approximation}
\author{Jie Liu, Li-Bin Fu}

\begin{abstract}
Adiabatic approximation for quantum evolution is investigated
quantitatively with addressing its dependence on the Berry
connections. We find that, in the adiabatic limit, the adiabatic
fidelity may uniformly converge to unit or diverge  manifesting the
breakdown of adiabatic approximation, depending on the type of the
singularity of the Berry connections as the functions of
slowly-varying parameter $R$. When  the Berry connections have a
singularity of $1/R^\sigma$ type with $\sigma < 1$, the adiabatic
fidelity converges to unit in a power-law; whereas when the
singularity index $\sigma$ is larger than one, adiabatic
approximation breaks down.
 Two-level models are used to
substantiate our theory.
\end{abstract}

\affiliation{Theoretical Physics Division, Institute of Applied
Physics and Computational Mathematics, P.O.Box 8009 , Beijing
100088, People's Republic of China} 
\maketitle

The adiabatic theorem, as a fundamental theorem in quantum
mechanics, plays a crucial role in our understand and manipulation
of microscopic world\cite {rev1}. Recent years witness its growing
importance in the context of quantum control, for example,
concerning adiabatic transportation of the newly formed matter ---
Bose-Einstein condensate\cite{liu1}, as well as for adiabatic
quantum computation\cite{fu}.

 Despite the
existence of an extensive literature\cite{addadd} on proofs of
estimates needed to justify the adiabatic approximation, doubts have
been raised about its validity leading to confusion about the
precise conditions needed to use it \cite{ref1,ref2}. In the present
paper, we point out that the above confusions can be avoided with
formulating the quantum  adiabatic approximation within
 parameter domain rather than  time domain. Within this
formulation, we address the fidelity of the adiabatic approximation
quantitatively (up to the prefactors of power-law or exponential
forms). In particular, we find that in the adiabatic limit, the
behavior of adiabatic fidelity depends on the singularity of the
Berry connections. That is, the Berry connections of the system
 determine how good the adiabatic approximation is. As we know, in
adiabatic quantum search algorithms, the upper bound of adiabatic
fidelity (measure of distance between exactly solution and adiabatic
approximation) is essential to the search time \cite{fidelity,wys}.

 The
system we consider is a
Hamiltonian containing slowly-varying dimensionless parameters $\mathbf{R}%
(t) $ belongs to a given regime $[\mathbf{R}_0, \mathbf{R}_1]$, saying, $H(%
\mathbf{R}(t))$. Initially we have a state, for example, the ground state $%
|E_{0}(\mathbf{R}(t_0))\rangle $ with energy $E_{0}(\mathbf{R}(t_0))$. The
wave function $|\Psi (t)\rangle $ fulfills the usual Schr\"{o}dinger
equation, i.e., $i\frac{d\Psi}{dt}=H(\mathbf{R}(t))\Psi (t)$, with $\hbar=1$%
. The above problem has a well-known adiabatic approximate solution ,
\begin{equation}
|\Psi_{ad}> = e^{-i\int^{t}E_{0}dt}e^{i\gamma _{0}}|E_{0}(\mathbf{R}(t))>.
\end{equation}
with $\gamma _{0}=i\int^{t}dt<E_{0}|\dot{E}_{0}>,$ the geometric phase term
\cite{berry, liu}. The above equation is the explicit formulation of the
adiabatic theorem stating that the initial nondegenerate ground state will
remain to be the instantaneous ground state and evolve only in its phase,
given by the time integral of the eigenenergy (known as the dynamical phase)
and a quantity independent of the time duration (known as the geometric
phase).

The problems is, how close the above adiabatic approximate solution to the
real solution $|\Psi (t)\rangle $. To clarify the above question and
formulate it quantitatively, we introduce two physical quantities, namely,
adiabatic parameter and adiabatic fidelity, respectively.

The dimensionless adiabatic parameter is defined as the ratio between the
change rate of the external parameters and the internal characteristic time
scales of the quantum system (i.e., the Rabi frequency $|E_{m}-E_{n}|$),
used to measure how slow the external parameter changes with respect to
time,
\begin{equation}
\epsilon = max \frac{|\mathbf{\dot{R}}|}{|E_{n}(\mathbf{R})-E_{m}(\mathbf{R})|%
} ,m\neq n.
\end{equation}
$\epsilon \to 0$ corresponds to adiabatic limit.
%%%%%%%%%%%%%%%%%%%%%%%%%%%%%%%%%%%%%%%%%%%%%%%%%%%%%%%%%%%%%%%%%%%%%%%%%%%%%%%%%%%%%%%%%
\begin{figure}[!b]
\begin{center}
\rotatebox{0}{\resizebox *{8.5cm}{4.0cm} {\includegraphics
{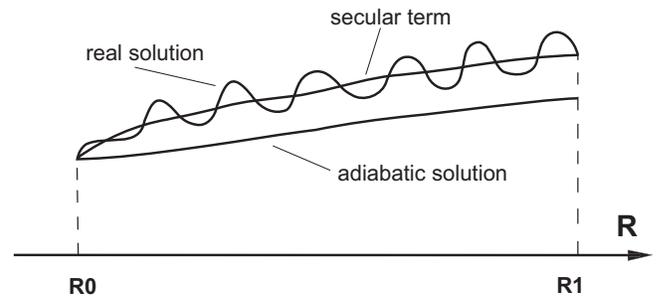}}}
\end{center}
\caption{Schematic illustration of the quantum adiabatic evolution
formulated in parameter condition, see text for detailed description.}
\label{fig1}
\end{figure}
%%%%%%%%%%%%%%%%%%%%%%%%%%%%%%%%%%%%%%%%%%%%%%%%%%%%%%%%%%%%%%%%%%%%%%%%%%%%%%%%%%%%%%%%%%

Adiabatic fidelity is introduced to measure how close the adiabatic solution
to real one, $F_{ad}=|\langle \Psi(t) |\Psi_{ad}\rangle |^{2}$. The
convergence of the the adiabatic fidelity to unit uniformly in the range $%
\mathbf{R}\in \lbrack \mathbf{R}_0 ,\mathbf{R}_1] $ in the adiabatic limit ($%
\epsilon \to 0$), indicates the validity of the adiabatic
approximation. Evaluating fidelity function will give an
estimation on how good  the adiabatic approximation is.

In Fig.1 we schematically illustrate the physical process we describe above.
Our main result is that, the distance between the adiabatic solution and
real one consists two parts: the fast oscillation term and the secular term.
The time scale of the oscillation is the Rabi period, its amplitude is
proportional to the square of the adiabatic parameter. The amplitude of the
secular term is exponential small ($\sim \exp{-1/\epsilon}$) suppose that
the Berry connections of the system are regular, and turn to be power-law ($%
\sim \epsilon^x, x<2$) if the Berry connections have singularity or the
external parameters vary in time nonlinearly.

We start our statement with writing the wavefunction as a superposition of
the instantaneous eigenstates, $|\Psi (t)\rangle
=\sum_{n}C_{n}(t)e^{-i\int^{t}dt(E_{n}-i\langle E_{n}(R)|\dot{E}%
_{n}(R)\rangle )}|E_{n}(\mathbf{R}(t))\rangle $, and suppose
initial state is the ground state, i.e., $C_0(t=0)=1,C_n(t=0)=0
,(n\ne 0)$. Then the adiabatic approximate solution takes form of
Eq.(1) and  the adiabatic fidelity $F_{ad}=|\langle \Psi(t)
|\Psi_{ad}\rangle |^{2}=|C_0|^2 \sim 1-|\Delta C_n|^2, (n\ne 0)$ .
To evaluate the adiabatic fidelity we need quantitatively evaluate
the change of the coefficients $C_{n}$ with respect to time.

Substituting the above solution  into Schr\"{o}dinger equation, we
have following differential equation for the coefficients,
\begin{equation}
\frac{d}{dt}C_{n}=i\sum_{m\neq n}e^{i\int^{t}((E_{n}-\mathbf{\alpha }_{nn}%
\mathbf{\dot{R}})-(E_{m}-\mathbf{\alpha }_{mm}\mathbf{\dot{R}}))dt}\mathbf{%
\alpha }_{nm}(\mathbf{R})\frac{d\mathbf{R}}{dt}C_{m}.
\end{equation}
where $\alpha _{nm}(\mathbf{R})$ is the Berry connection. Both off-diagonal
and diagonal Berry connections have clear physical meaning and important
applications\cite{offdia}. We first suppose these Berry connections and the
gradient of instantaneous eigenenergy are not singular (NS) as the functions
of the external parameters, i.e.,
\begin{equation}
\mathbf{\alpha }_{nm}(\mathbf{R})=\langle E_{n}|i\nabla _{\mathbf{R}%
}|E_{m}\rangle ;\mathbf{\beta }_{n}(\mathbf{R})=\nabla _{\mathbf{R}}E_{n}(%
\mathbf{R}),\mathit{NS}.
\end{equation}

%Under the above condition and in the adiabatic limit,
%the exponent in Eq.(3)
%can be approximated as $e^{i(E_{n}-E_{m})t+i\int^{t}((\beta_{n}-\mathbf{%
%\alpha }_{nn} )\mathbf{\dot{R}}-(\beta_{m}-\mathbf{\alpha }_{mm})\mathbf{%
%\dot{R}})dt} \approx e^{i(E_{n}-E_{m})t}$.
%%%%%%%%%%%%%%%%%%%%%%%%%%%%%%%%%%%%%%%%%%%%%%%%%%%%%%%%%%%%%%%%%%%%%%%%%%%%%%%%%%%%%%%%%
\begin{figure}[!b]
\begin{center}
\rotatebox{0}{\resizebox *{8.5cm}{4.0cm} {\includegraphics
{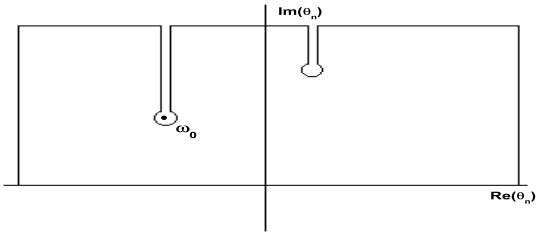}}}
\end{center}
\caption{The integral path and the singular points in the complex
plane.} \label{fig2}
\end{figure}
%%%%%%%%%%%%%%%%%%%%%%%%%%%%%%%%%%%%%%%%%%%%%%%%%%%%%%%%%%%%%%%%%%%%%%%%%%%%%%%%%%%%%%%%%%
Under the above condition and in the adiabatic limit, to estimate
the distance between the adiabatic solution and the real one, we
need quantitatively evaluate the change of the coefficients
$C_{n}$ with respect to time.
 The right-hand of Eq.(3) contains unknown  $C_m$, to the first order of
  approximation, we take $C_0=1$ and $C_m=0, m\neq 0$ in the right-hand of Eq.(3).
  Then,
 Eq.(3) shows that, the
change consists two parts: the fast oscillation term and the
secular term. The time scale of the oscillation is the Rabi
period, its amplitude is proportional to the adiabatic parameter
suppose that the Berry connections are regular with limitation.
Whereas the  secular term maybe is exponential small of form
$e^{-\frac{1}{\epsilon}}$ or power-law depending on the Berry
connections in following.

Let us denote $\theta _{n}=\int^{t}(E_{n}-E_{0})dt$, and then the
upper bound of the increment on the coefficients ($n\neq 0$) can
be evaluated as follows
\begin{eqnarray}
\Delta C_{n} &\sim &\int_{\theta _{n}(\mathbf{R}_{0})}^{\theta _{n}(\mathbf{R%
}_{1})}\frac{e^{i\theta _{n}}}{E_{n}-E_{0}}\mathbf{\alpha }_{n0}\dot{\mathbf{R}}%
d\theta _{n} \\
&=&\int_{\mathbf{-\infty }}^{\mathbf{\infty }}...d\theta _{n}\mathbf{-}%
\left( \int_{\mathbf{-\infty }}^{\theta_n(\mathbf{R}_{0})}\mathbf{+}\int_{\theta_n(\mathbf{R}%
_{1})}^{\mathbf{\infty }}\right) ...d\theta_n
\end{eqnarray}
where we set that in the right-hand of Eq.(3) the coefficients
$C_{m}\sim 0$ for $m\neq 0,$ and $C_{0}\sim 1$ since we want to
estimate the upper bound of the adiabatic approximation. For
simple and without losing generalization, in following deductions
we regards the slowly-varying as a scalar quantity, and assume
that $\frac{dR}{dt} \sim \epsilon g(R)$ with function $g(R)$ being
regular with limitation.

The main contribution to the first term on the right hand comes
from the pole
point, $\theta _{n}^{c}=\int^{t_c}(E_{n}-E_{0})dt \sim \frac{1}{\epsilon}\int^{R_c}(E_{n}-E_{0})/g(R)dR$,
determined by the equation $E_{n}(\mathbf{\theta^c }%
_{n})-E_{0}(\mathbf{\theta^c }_{n})=0$. Because the non-degeneracy
of the system, the solutions of the above pole points are complex
with imaginary parts. Let $\omega_0$ be the singularity nearest
the real axis, i.e., the one with the smallest (positive)
imaginary part (see Fig.2). Then, the first term is approximately
bounded by $exp(-Im \omega_0) \sim e^{-\frac
{1}{\epsilon}|\int^{|Im(R_c)|}(E_n-E_0)/g(R)dR|}$, which is just
the secular term and is exponential small in adiabatic
limit\cite{liuliu}. The second integral is proportional to
$\epsilon$, which gives the estimation on  the fast oscillation
term.

Now, we consider the situation that Berry connection has
singularity in the parameter regime $[\mathbf{R}_0,
\mathbf{R}_1]$, e.g., $R=0$, taking the form $\frac{1}{R^{\sigma
}}$. We then evaluate the above  integral  in the neighborhood
domain $[-\epsilon, \epsilon]$ of the singular point, other regime
is regular and can be treated in the same way as before. Then,
\begin{equation}
|\Delta C_n| \sim |\int_{-\epsilon}^{\epsilon} e^{i\int^{t}(E_{n}-E_{0})dt}%
\mathbf{\alpha }_{n0}(R)dR|\sim \epsilon^{(1-\sigma )}.
\end{equation}
The situation is divided into two cases: $\sigma <1$ and $\sigma \ge 1$. For
$\sigma <1$, this type of singularity can be removed because the integral is
finite. The integral in the neighborhood domain $[-\epsilon, \epsilon]$ of
the singularity contributes a quantity of order $\epsilon^{(1-\sigma )}$, We
thus expect that the adiabatic fidelity approaches to unit uniformly in the $%
2(1-\sigma )$ power-law of the adiabatic parameter, i.e.,
\begin{equation}
1-F_{ad} \sim \epsilon^{2(1-\sigma)}.
\end{equation}
For the case of $\sigma \geq 1$, the singularity is irremovable
and the adiabatic approximation is expected to break down.

The above discussions is readily extended to the case
 that the
slowing-varying parameters change nonlinearly with time, i.e.,
$R=\epsilon t^\sigma$, with $\sigma$ is any positive number. The
nonlinear time dependent parameter has many physical origins, for
example in molecule spin system the effective field vary in time
nonlinearly\cite{nonlinear1}. Another field of broad examples  is
quantum optics, Rabi frequency coupling different levels (i.e.,
stimulated Raman adiabatic passage)  is usually nonlinear dependence
on time \cite{nonlinear2}.
 Here we suppose that the Berry connections
of quantum system are regular with limitation as the function of
the parameter $R$ and the level spacings are of order 1. To apply
our theory, we introduce the new parameters $R'$ and $\epsilon'$
through the expressions $\epsilon'=\epsilon^{\frac{1}{\sigma}} $
and $R'=\epsilon' t$. Then, $R={R'}^\sigma$. As a function of the
new parameter $R'$, the singularity of the Berry connections
determined by the $\frac{dR}{dR'} \sim 1/{R'}^{1-\sigma}$. Our
discussions are divided into two cases: i) $\sigma > 1$ and ii)
$\sigma < 1$. In the former case, the Berry connections as the
functions of the new parameter are regular,  so the adiabatic
fidelity is determined by the short-term oscillation and is
expected to converge to one in a power-law of $\epsilon'^2$. Then
we have
\begin{equation}
1-F_{ad} \sim \epsilon ^{\frac{2}{\sigma}}.
\end{equation}
In the latter case, the Berry connections as the functions of the
new parameter are singular of type $1/{R'}^{1-\sigma}$.
Fortunately, this singularity is removable,  it gives an upper
bound of the adiabatic Fidelity as $\epsilon'^{2\sigma}$, i.e.,
\begin{equation}
1-F_{ad} \sim \epsilon^2.
\end{equation}
Notice that in this case, the upper bound of the adiabatic
fidelity is independence of the nonlinear index $\sigma$ .

In following, we use two-level models to substantiate the above discussions.
Our model is denoted by $S^{a}$: a spin-half particle in a rotating magnetic
field, its Hamiltonian reads,
\begin{eqnarray}
H^{a}(R(t)) &=&-\frac{\omega _{0}}{2}(\sin \theta \cos f(R(t))\sigma
_{x}+\sin \theta \sin f(R(t))\sigma _{y}  \notag \\
&&+\cos \theta \sigma _{z}),  \label{sa}
\end{eqnarray}
$\omega _{0}$ is defined by the strength of the magnetic field, and $\sigma
_{i},$ $(i=x,y,z)$ are Pauli matrices. $f(R)$ is a function of
slowly-varying parameter $R.$ The above $2\times 2$ matrix is readily
diagonalized for fixed $R$ and we then obtain its instantaneous \
eigenenergies $E_{\pm }^{a}=\pm \frac{\omega _{0}}{2}$, and instantaneous
eigenvectors,
\begin{equation}
|E_{-}^{a}(R)\rangle =\left(
\begin{array}{c}
e^{-i\frac{f(R)}{2}}\sin \frac{\theta }{2} \\
-e^{i\frac{f(R)}{2}}\cos \frac{\theta }{2}
\end{array}
\right) ,|E_{+}^{a}(R)\rangle =\left(
\begin{array}{c}
e^{-i\frac{f(R)}{2}}\cos \frac{\theta }{2} \\
e^{i\frac{f(R)}{2}}\sin \frac{\theta }{2}
\end{array}
\right)
\end{equation}

The Berry connections are derived as follows,
\begin{equation}
\alpha^c _{--}=\frac{1}{2}\cos \theta \frac{df}{dR},\;\alpha^c _{-+}=\frac{1%
}{2}\sin \theta \frac{df}{dR}.  \label{ccc}
\end{equation}
%%%%%%%%%%%%%%%%%%%%%%%%%%%%%%%%%%%%%%%%%%%%%%%%%%%%%%%%%%%%%%%%%%%%%%%%%%%%%%%%%%%%%%%%%
\begin{figure}[tbh]
\begin{center}
\rotatebox{0}{\resizebox *{8.5cm}{7.0cm} {\includegraphics
{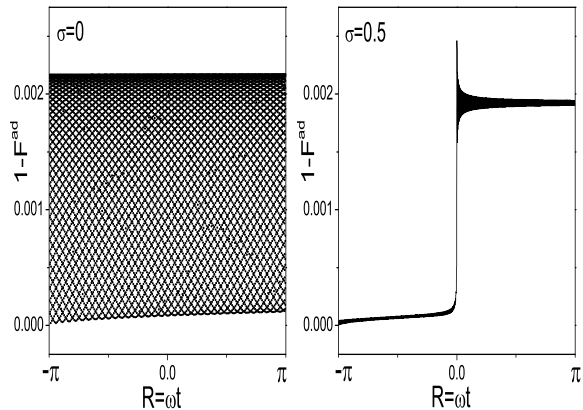}}}
\end{center}
\caption{Adiabatic fidelity evolves for different type of
singularity. } \label{fig3}
\end{figure}
%%%%%%%%%%%%%%%%%%%%%%%%%%%%%%%%%%%%%%%%%%%%%%%%%%%%%%%%%%%%%%%%%%%%%%%%%%%%%%%%%%%%%%%%%%
%%%%%%%%%%%%%%%%%%%%%%%%%%%%%%%%%%%%%%%%%%%%%%%%%%%%%%%%%%%%%%%%%%%%%%%%%%%%%%%%%%%%%%%%%
\begin{figure}[tbh]
\begin{center}
\rotatebox{0}{\resizebox *{8.5cm}{7.0cm} {\includegraphics
{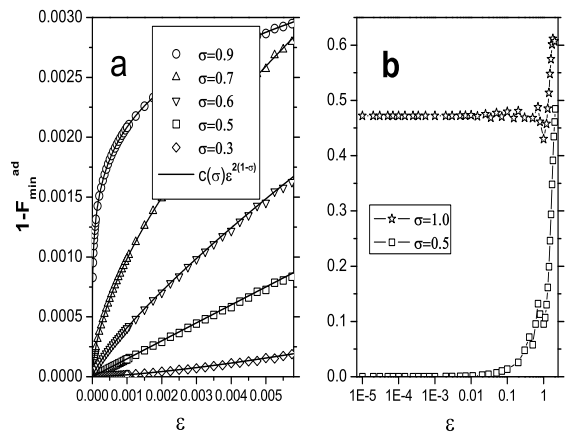}}}
\end{center}
\caption{Upper bound of the adiabatic fidelity in system $S^{c}$
for different type of singularity. $F^{ad}_{min}$ is the minimum
fidelity in the parameter range $\mathbf{R}\in \lbrack
-2\protect\pi,2\protect\pi] $} \label{fig4}
\end{figure}
%%%%%%%%%%%%%%%%%%%%%%%%%%%%%%%%%%%%%%%%%%%%%%%%%%%%%%%%%%%%%%%%%%%%%%%%%%%%%%%%%%%%%%%%%%
i)We first consider following two cases: $f(R)=\ln |R|$ and $%
f(R)=|R|^{1-\sigma }$ with $\sigma <1$, respectively. And here the
slowly-varying parameter is supposed to linearly change with time, i.e., $%
R(t)=\omega t$. $\omega $ is the rotating frequency of the
magnetic field. Apparently, the Berry connection has singularity
of the form $1/R^{\sigma }$ at point $R=0$ . These systems are
complicated and analytic solutions are not reachable. We thus make
numerical simulations on the adiabatic fidelity by directly
solving the Schr\"{o}dinger equation with the $4^{th}-5^{th}$
Runge-Kutta adaptive step method. Our results are shown in Fig.3
and Fig.4. In the Fig.3, it is clearly shown that, without the
singularity (Fig.3a), the distance between the adiabatic solution
and real solution is determined by the fast oscillation, therefore
gives the upper bound of square adiabatic parameter. With the
singularity (Fig.3b), the upper bound is determined by the type of
the singularity of the Berry connections as we discuss above. For
the situation that the Berry connections have irremovable
singularity (i.e., $\sigma =1$, see Fig.4b) the adiabatic fidelity
converges to 0.53 rather than one implying the failure of
adiabatic approximation. For the case of the removable singulary
of the Berry connection($\sigma <1$) the adiabatic fidelity
converges to unit in the power-law dependence of the adiabatic
parameters as we expect (see Fig.4a).

ii) We then set $f(R)=R$ and $R$ varies in time nonlinearly, i.e., $%
R=\epsilon sign(t) |t|^\sigma$. In this case, we see that (Fig.5), for $%
\sigma >1$, the adiabatic fidelity converges to unit in a power-law of
exponent $2/\sigma$; for $\sigma < 1$, it clearly demonstrates that the
exponent turns to be two, independence of the nonlinear index $\sigma$.
These numerical simulations corroborate our theory.
%%%%%%%%%%%%%%%%%%%%%%%%%%%%%%%%%%%%%%%%%%%%%%%%%%%%%%%%%%%%%%%%%%%%%%%%%%%%%%%%%%%%%%%%%
\begin{figure}[tbh]
\begin{center}
\rotatebox{0}{\resizebox *{8.5cm}{7.0cm} {\includegraphics
{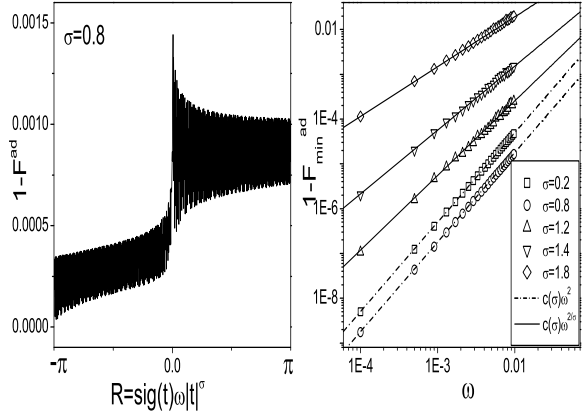}}}
\end{center}
\caption{Upper bound of the adiabatic fidelity for different type
of the nonlinearly-varying external parameters.} \label{fig5}
\end{figure}
%%%%%%%%%%%%%%%%%%%%%%%%%%%%%%%%%%%%%%%%%%%%%%%%%%%%%%%%%%%%%%%%%%%%%%%%%%%%%%%%%%%%%%%%%%

In the above discussion we discuss quantum adiabatic issue under
the parameter conditions. We emphasize that the adiabatic problem
can only be well formulated  in the parameter domain but not in
time domain. Therefore, it requires that the Hamiltonian depends
on the time only through the slowly-varying parameters (i.e.,
taking form $H(R(t))$) and, the range  of the parameters (i.e.,
($R_0$,$R_1$))  can be reached at certain time (i.e.,$t_0$ and
$t_1$) for any small adiabatic parameter ($\epsilon$) as sketched
in Fig.1. Usually, the smaller the adiabatic parameter, the longer
the time duration (i.e., $t_1-t_0$) is needed. In following we
address this point with an example raised in \cite{ref2}.  It is
constructed from $S^{a}$ for the case of $f(R)=R, R=\omega t$
through following relation,
\begin{equation}
H^{ce} = -U^{a\dagger }(t)H^{a}(t)U^{a}(t)
\end{equation}
with $U^{a}(t)=T\exp (\int_{0}^{t}H^{a}(t^{\prime })dt^{\prime })$ the time
evolution operator of system $S^{a}$. Its explicit analytic expression is
readily obtained \cite{ref2}. After lengthy deduction, we obtain the
explicit expression of the Hamiltonian $H^{ce} = \frac{\omega _{0}}{2}%
\mathbf{L}(t)\cdot \mathbf{\sigma},$ where $\mathbf{L}(t)=(\sin
\theta (\omega _{0}^{2}+2\omega \omega _{0}\cos \theta \cos
^{2}\varpi t/2+\omega ^{2}\cos \varpi t)/\varpi ^{2},\frac{\omega
\sin \theta }{\varpi }\sin
\varpi t,$ $\cos \theta +\frac{2\omega \omega _{0}\sin \theta }{\varpi ^{2}}%
\sin ^{2}\theta \sin ^{2}\varpi t/2),$ and $\varpi =\sqrt{\omega
_{0}^{2}+\omega ^{2}+2\omega \omega _{0}\cos \theta }$.

It is easy to verify that $\mathbf{L}(t)$ is a unit vector, i.e, $|\mathbf{L}%
(t)|=1.$ The eigenvalues and eigenvectors for this system are,
\begin{equation}
E_{\pm }^{ce}=\pm \frac{\omega _{0}}{2},\;\;\left| E_{\pm }^{ce}(\mathbf{L}%
)\right\rangle =\left(
\begin{array}{c}
\sqrt{\frac{1\pm L_{3}}{2}}e^{-i\phi } \\
\mp \sqrt{\frac{1\mp L_{3}}{2}}e^{i\phi }
\end{array}
\right) ,
\end{equation}
where $\phi =\frac{1}{2}\arctan (L_{2}/L_{1}).$ We then can obtain
the Berry connections as follows,
\begin{eqnarray}
\mathbf{\alpha }_{--}^{ce} &=&(\frac{L_{2}L_{3}}{2(1-L_{3}^{2})},-\frac{%
L_{1}L_{3}}{2(1-L_{3}^{2})},0),  \notag \\
\mathbf{\alpha }_{-+}^{ce} &=&(-\frac{L_{2}}{2\sqrt{1-L_{3}^{2}}},\frac{L_{1}%
}{2\sqrt{1-L_{3}^{2}}},-\frac{i}{2\sqrt{1-L_{3}^{2}}}).
\label{bcc}
\end{eqnarray}
As $L_{3}<1$, the Berry connections are not singular. For $d|\mathbf{L|}%
/dt\sim \omega $, the adiabatic parameter of this system is
$\epsilon =\omega /\omega _{0}.$

The controversy is that, even though in the adiabatic limit
$\epsilon \to 0$, the adiabatic fidelity calculated in the time
domain $t \in [-2\pi, 2\pi]$ does not diverge to unit\cite{ref2}.
Moreover, with changing  the sign of the above Hamiltonian and
re-calculating the adiabatic fidelity in the time domain $[-2\pi,
2\pi]$,  we find that the adiabatic fidelity converges to unit in
the adiabatic limit. The above result is rather confused. The
reason for the  above controversy is that the problem is discussed
in time domain rather in parameter domain.

To resolve the above controversy, we check the above system in the
parameter domain . First, after transformation (14), the $R=\omega
t$ acted as the slowly-varying parameter in $H^a(t)$ system no
longer should be chosen as  the slowly-varying parameter of the
new system $H^{ce}$, because the Hamiltonian $H^{ce}$ depends
explicitly on the time not only through $R$.  Instead,
$\mathbf{L}(t)$ can serve as the slowly-varying parameters.
However, the range of the parameters keeps the same order of the
adiabatic parameter (i.e.,
$|\mathbf{L_1}(t=t_1)-\mathbf{L_0}(t=t_0)| \sim \omega$)  and
tends to zero in the adiabatic limit no matter how long the
evolution time $t_1-t_0$ is.  This   completely counters to our
picture schematically plotted in Fig.1. Above analysis indicates
that the system $H_{ce}$ can not be well formulated in the
parameter domain, it essentially not a system that adiabatic
theory can applies to. If one  discuss the dynamics of this system
in the time domain as shown the above, any strange things can
happen.

In summary, we investigate the fidelity for quantum evolution under
the parameter domain with addressing the adiabatic approximation
quantitatively. Within this framework, we clarify the confusions in
applying quantum adiabatic theory, and  find that the singularity of
Berry connections  inhibit the accuracy of the adiabatic
approximation. Our estimation on the adiabatic fidelity  has
important meaning in the practical adiabatic quantum search
algorithms .

This work was supported by National Natural Science Foundation of
China (No.10474008,10604009), Science and Technology fund of CAEP,
the National Fundamental Research Programme of China under Grant
No. 2005CB3724503, the National High Technology Research and
Development Program of China (863 Program) international
cooperation program under Grant No.2004AA1Z1220.

\end{document}